# Method for removing interference in chaotic signals based on the Lyapunov exponent

YANG Nan[1], LONG Zhang-Cai[1] [*], ZHAO Xiang-Hui[1]

[1]*College of Physics, Huazhong University of Science and Technology, Wuhan 430074, PR China*

A new method based on the phenomenon of synchronization and the properties of chaos is proposed to reduce interference in the transferred chaotic signals of synchronized systems. In this paper, the interference is considered as a series of small deviations from the original clean trajectory in the phase space. By means of our special design, these small deviations can be estimated using positive Lyapunov exponents, and removed from interfered chaotic signals. Application is illustrated for the Lorenz attractor, and numerical computing demonstrates that the method is effective in removing typical external interference.

**noise reduction, chaotic synchronization, Lyapunov exponent**

Chaotic synchronization [1] and its potential applications [2–12] have attracted increasing interest during the past two decades. One of the fundamental problems arising in the study of chaotic synchronization is that the transferred chaotic signals are masked by environmental noise [13]. Traditional signal processing methods based on spectrum analysis are unable to remove noise from chaotic signals without distorting the signals. Many approaches have been proposed to smooth the chaotic trajectories or extract the signal component from noisy experimental chaotic data [14–24]. Some of them, such as the Local Geometric Projection Method [18], have been successfully put into practice. However, based on the characters of noise—irregular and nonsmooth, almost all the previous noise-reducing methods for chaotic signals are only robust in the cases of white noise or high-dimensional noise. In a natural environment, however, the transferred chaotic signals may suffer from various types of interference, including not only the introduction of stochastic noise but also unpredictable regular signals. Therefore, although the problem of noise reduction for chaotic signals has been widely studied in the literature, further research is necessary [13].

In the present work, we propose a method to estimate and reduce unknown interference in transferred chaotic signals, which could distinguish a chaotic signal from external interference including regular interference. Instead of using the properties of the interference signal or noise, our method resorts to the nonlinear statistical characteristic of the chaotic signal, the positive Lyapunov exponent, which can be estimated from the long-time average divergence rate of small

Corresponding author：longzc01@mails.tsinghua.edu.cn
This work is supported by the National Natural Science Foundation of China (Grant No 60874099) and "Cognitive Computing of Visual and Auditory Information" of National Natural Science Foundation of China (Grant No 90820001).

deviations between adjacent orbits, as presented by Wolf et al. [31]. As the reverse process to Wolf's method of estimating Lyapunov exponents, in this study, we estimate the unknown initial deviations from foreknown positive Lyapunov exponents. While the external interference is considered a series of small unknown deviations in chaotic signals, by means of our special design, these small deviations can evolve freely in an autonomous chaotic system and be statistically estimated from the positive Lyapunov exponents of this autonomous chaotic system.

## 1. Algorithm of interference reduction

To simplify the model, we only consider the completely synchronized cases of three-dimensional chaotic systems. We focus on a couple of completely synchronized chaotic systems: $X$ ($d\mathbf{x}/dt = \mathbf{f}(\mathbf{x}, s(t))$) and $Y$ ($d\mathbf{y}/dt = \mathbf{f}(\mathbf{y}, s'(t))$). The drive system $X$ provides the chaotic signal $s(t)$, and the response system $Y$ provides synchronous responses to the output of $X$. Nevertheless, the signal received by the response system $Y$ is contaminated by external noise $n(t)$, which can be represented as $s'(t) = n(t) + s(t)$. (All noise in our research is additive noise, which means that we simply add it linearly to the clean signal data.) Our goal is to remove interference $n(t)$ from the received signal $s'(t)$.

We consider the interference as deviations from the original orbit, and this deviated state evolves freely in an auxiliary autonomous chaotic system, which is designed as a copy of the drive system. Therefore, the chaotic orbit in the autonomous system would drastically differ from the original orbit. When the evolving time $T$, the deviation distance $L$ and the average divergence rate $\lambda_{max}$ (the positive Lyapunov exponent) of the autonomous chaotic system are known, the initial deviation can be estimated and then removed from the original noisy orbit. The scheme is shown in Fig. 1, in which the noisy chaotic signals in the response system $Y$, $\mathbf{y} = [\mathbf{y}(t_1), \mathbf{y}(t_2), ..., \mathbf{y}(t_N)]$ (solid curve with symbols), follow the clean orbit of drive signals $X$, $\mathbf{x} = [\mathbf{x}(t_1), \mathbf{x}(t_2), ..., \mathbf{x}(t_N)]$ (dotted curve). Meanwhile, although having the same noisy initial condition (asterisk symbols) with the orbit of system $Y$, the orbit of autonomous auxiliary system $Z$ (the solid line) evolves freely and departs from the original orbit (solid curve with symbols): $\mathbf{z}_i = [\mathbf{z}_i(t_1), \mathbf{z}_i(t_2), ..., \mathbf{z}_i(t_N)]$. $u$ denotes the interference to be removed, and $u'$ is the estimation of $u$.

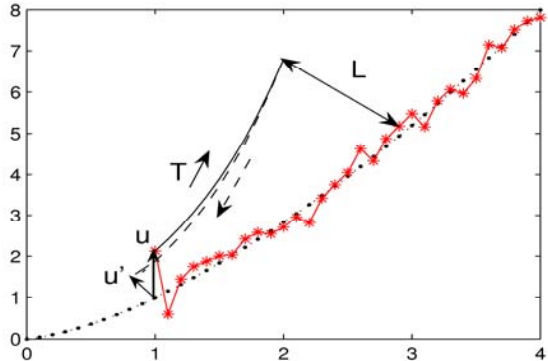

Figure 1. Schematic illustration of noise estimation. The solid line with symbols denotes the noisy response trajectory $\mathbf{y}(t)$, while the dotted black line represents the clean trajectory $\mathbf{x}(t)$. u denotes the noise point to be reduced (which is assumed to be unknown), and u' is the estimation of u. $L$ is the fixed divergence length and $T$ represents the evolving time from u to $L$.

When the evolving time $\tau_i$ from the small deviation $u(t_i)$ to a fixed length $L$ is known, the estimated noise $u'(t_i)$ can be computed by

$$\|u'(t_i)\| = L\exp[-\lambda_{\max}(\tau_i)], \quad i=1,2,...,N. \quad (1)$$

Here $\lambda_{\max}$ denotes the average divergence rate (the positive Lyapunov exponent). Although there is the influence of external noise, the slaved trajectory $\mathbf{y}(t)$ does not depart from $\mathbf{x}(t)$. The errors between $\mathbf{x}(t)$ and $\mathbf{y}(t)$ are sufficiently small compared with $L$ because the additive noise does not destroy the state of synchronization [19]. Thus, the fixed length $L$ is approximately presented as

$$L \approx \|\mathbf{z}_i(t_0+\tau_i) - \mathbf{y}(t_i+\tau_i)\|, \quad i=1,2,...,N. \quad (2)$$

To ensure that the estimated noise $u'(t)$ is less than realistic noise $u(t)$, the parameter $\eta$ is added to $\tau_i$ in Eq. (1), and the estimated interference is

$$\|u'(t_i)\| = L\exp[-\lambda_{\max}(\tau_i+\eta)], \quad i=1,2,...,N. \quad (3)$$

The noisy points $\mathbf{y}(t_1)$, $\mathbf{y}(t_1)$,…, $\mathbf{y}(t_N)$ then correspond to $N$ freely evolving orbits $\mathbf{z}_1(t), \mathbf{z}_2(t), ..., \mathbf{z}_n(t)$. This process is repeated until each noise point is estimated. We make an approximation that the interference $u(t)$ existing in transferred signals $s(t)$ is equal to the errors between the trajectories $\mathbf{x}(t)$ and $\mathbf{y}(t)$, $\|n(t)\| \approx \|\mathbf{x}(t) - \mathbf{y}(t)\|$. Finally, we remove the estimated noise $u'(t)$ from $n(t)$. Thus far, our method is feasible.

The interference is signed data and defined as

$$\mathrm{sign}(u(t_i)) = \begin{cases} 1 & \text{if } u(t_i) >= 0 \\ -1 & \text{else } u(t_i) < 0 \end{cases}. \quad (4)$$
$$i=1,2,...,N$$

Combining functions (3) and (4), the estimated noise with signs is given by

$$u'(t_i) = \mathrm{sign}[u(t_i)]L\exp[-\lambda_{\max}(\tau_i+\eta)], \quad i=1,2,...,N. \quad (5)$$

We can apply the method to reduce the remaining noise repeatedly.

To avoid the case that the estimated noise is greater than the original noise, we only apply our method to the large interference data and leave the small data for the next reduction. The large and small interference data are distinguished in the following way. The transferred signals are considered linear in each local section. We compute the mean value for each section, which is treated as the boundary of large and small noise. We divide the time series $s'(t)$ into small sections $\mathbf{w}_1, \mathbf{w}_2, ..., \mathbf{w}_n$ with each section containing $k$ points: $\mathbf{w}_1 = [s'(t_1), s'(t_2), ..., s'(t_k)]$, $\mathbf{w}_2 = [s'(t_{k+1}), s'(t_{k+2}), ..., s'(t_{2k})]$ ... $\mathbf{w}_n = [s'(t_{nk-k+1}), s'(t_{nk-k+2}), ..., s'(t_{nk})]$. Here $k$ is a small number. For the $i$-th vector $\mathbf{w}_i$, we then calculate the mean value $\bar{\mathbf{w}}_i$ over all its elements. If $s'(t_j) < \bar{w}_i$, then $s'(t_j)$ is considered to be small noise and is not reduced; here $k(i-1)+1 \leq j \leq k(i-1)+k$.

## 2. Scheme for determining signs of noise points

A crucial hypothesis referred to in section 1 is that all the signs of noise points are known in advance. In this section, we present the scheme for determining signs in detail. The positive and negative noise points are absolutely indistinguishable when they are mixed in chaotic signals because the initial information of chaotic system is lost entirely in a short period. However, it makes sense for all the noise points to be negative or positive. In a word, our scheme to determine the signs of noise points is to simply add a direct current signal $d$ to the noise series $u(t)$. Thus, the new external noise is

$$v(t) = d + u(t). \qquad (6)$$

$v(t) > 0$ or $v(t) < 0$ is a necessary condition for the selection of $d$. Figure 2 illustrates the applied concept.

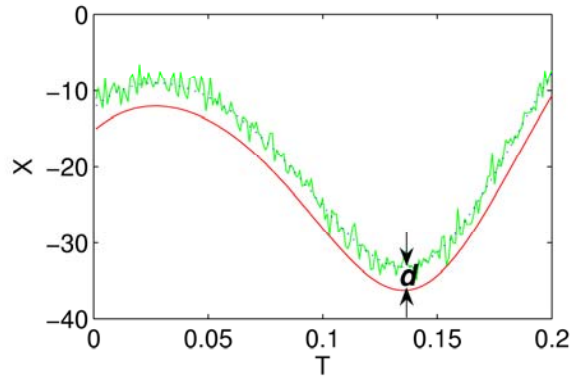

Figure 2. Adding a direct-current signal $d$ to a noisy transferred signal $s(t)$. The lower curve is the original clean transferred signal $s(t)$, which is treated as a criterion to judge the signs of noisy data. We add a direct-current signal $d$ to the noisy signal $s(t) + u(t)$. The upper solid curve is the translated noisy trajectory $s'(t) = s(t) + u(t) + d$.

The performance of this noise reduction method is limited by the selection of $d$. If $d$ is too large, extra computation must be carried out until the signals are sufficiently clean; otherwise, if the chosen $d$ is too small, there will be more orientation errors, which is fatal to the success of employing this method. To obtain the best-fit value of $d$, we first estimate the noise level. Several approaches for the estimation of noise levels have been proposed [26,27].

## 3. Numerical simulation

There are four steps to the numerical simulation for verifying the effectiveness of the proposed method.

1. A direct-current signal $d$ is added to received signals $s'(t)$ so that the signs of external noise $v(t) = u(t) + d$ are identical; i.e., $v(t) > 0$ or $v(t) < 0$.

2. The transferred signals $s'(t)$ are divided into small sections $\mathbf{w}_1, \mathbf{w}_2, ..., \mathbf{w}_n$, and the small noise is filtered out.

3. Each noise datum $v(t_i)$ is considered a small error in the chaotic system and conditions are established to have the error freely evolve to a fixed length $L$. Noise data are then estimated by introducing the largest (positive) Lyapunov exponent $\lambda_{max}$:

$$v'(t_i) = L\,\text{sign}[v(t_i)]\exp[-\lambda_{max}(\tau_i + \eta)], i = 1, 2, ..., N . \qquad (7)$$

4. Steps 2 and step 3 are repeated to obtain the best result.

We now apply this method to a couple of synchronizing Lorenz systems [28–30]. The equations of the signal generator are

$$dx/dt = \sigma(y - x) \qquad dy/dt = -xz + ry - y \qquad dz/dt = xy - bz . \qquad (8)$$

The equations of the receiving system are

$$dx'/dt = \sigma(y' - s(t)) \qquad dy'/dt = -s(t)z' + ry' - y' \qquad dz'/dt = s(t)y' - bz', \qquad (9)$$

where $\sigma = 10$, $r = 8/3$, and $b = 60$. The receiver is coupled with the generator only through the noisy transferred signal $s(t)$: $s(t) = x(t) + u(t)$, where $u(t)$ denotes the added noise. The time series $x(t)$ and $u(t)$ are sampled as $x(n) = x[t_0 + (n-1)\Delta t]$ and $u(n) = u[t_0 + (n-1)\Delta t]$, where $n = 1, 2, ..., N$.

We choose a sample distance $\Delta t = 10^{-3}$ seconds and time series $N = 5000$ points, with each section comprising 20 points. Here the dynamic models are assumed to be unknown. The positive Lyapunov exponent $\lambda = 1.40$, which is estimated from the transferred signals [31–34]. The parameters $\eta$, $d$, and $L$ are chosen as 1, 2, 6; here $L$ decreases by a factor of 1 after a step of noise removal. The signal to noise ratio (*SNR*) is defined as

$$SNR = 10\log_{10}(\sum_{i=1}^{N}[s(i)]^2 / \sum_{i=1}^{N}[u(i)]^2) . \qquad (11)$$

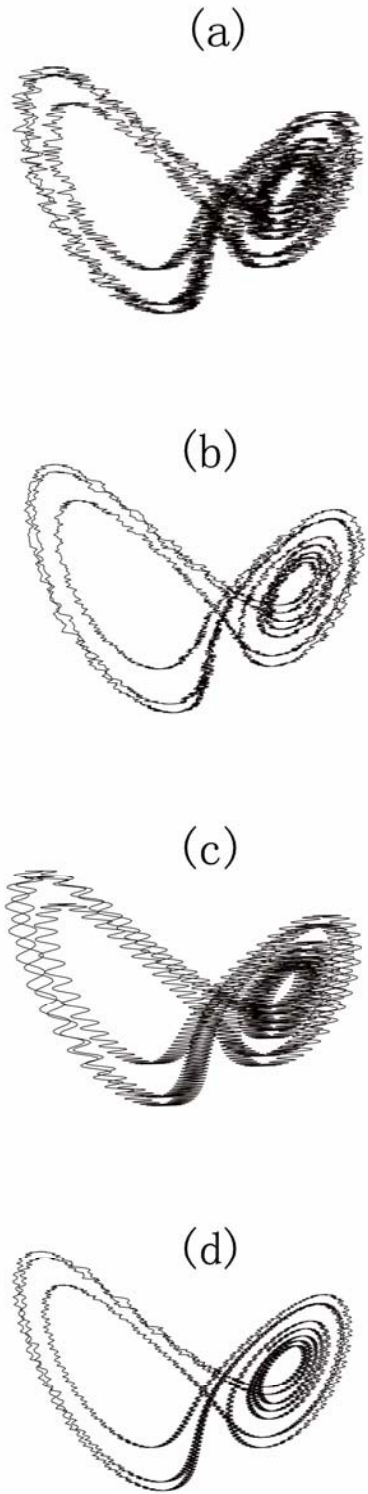

Figure 3. Noise reduction applied to Lorenz attractors. (a), (c) Before noise reduction in the cases of uniform white noise and sinusoidal interference added to transferred signals; (b), (d) the cases of (a) and (c) after noise reduction. Note that these diagrams are neither the phase portrait of the drive system nor that of the response system. They can be considered as phase portraits of the drive system with the output signals $\mathbf{x}$ replaced by the contaminated signals $\mathbf{x}+\mathbf{u}$.

Two types of interference—uniformly distributed white noise and a sine wave—are added to the transferred chaotic signals. Here, the white noise is considered random interference in the chaotic signal, while the sine wave is unknown regular interference. Panels (a) and (c) in Figure 3 are the phase portraits of Lorenz systems with these two types of interference added to transferred signals (x vector). Panels (b) and (d) present the corresponding chaotic data after noise reduction. *SNR* for panels (b) and (d) are: 26.1 and 27.3 dB, respectively, which are 6.4 and 9.3 dB higher than those for panels (a) and (c), respectively. We thus see significant improvements after removing noise in Figure 3.

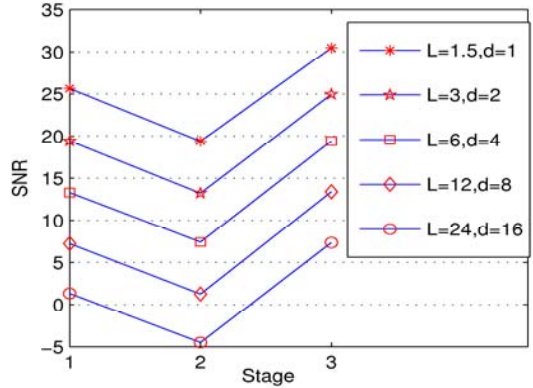

Figure 4. Performance of the noise reduction scheme for different noise levels. The horizontal axis denotes three stages of the noise reduction method. *SNR* is that of the transferred noisy signal in stage 1, that when the direct current *d* is added in stage 2, and that after removing noise five times in stage 3. The values of *d* and *L* increase with the increasing noise level.

We finally test our method for different levels of uniform white noise. The parameter $\eta=1$ in all cases. The values of direct current *d* and fixed length *L* increase as the noise level increases because the original errors need larger space to evolve when the noise level is higher. In Figure 4, although there is a sharp decrease when the direct current is added to the transferred signals, *SNR* rises rapidly as the noise reduction proceeds.

## 4. Conclusions and discussions

This paper proposed a new method for removing interference in chaotic synchronizing systems. Employing this method, the characteristic of synchronization and the positive Lyapunov exponent are used to estimate the interference. Encouraging results were obtained when the scheme was applied to typical chaotic systems such as a couple of synchronized Lorenz attractors. Finally, we investigated the effects of the noise level. The results indicate that this algorithm not only works well in low-noise cases but also is effective when the noise level is very high.

In a survey of current nonlinear noise reduction methods, we found that most are based on two principles: 1) the use of different correlations between noise and chaotic signals and 2) the smoothing of noisy trajectories in reconstructed space. These methods are effective for specific types of additive interference but are inapplicable to many other types of interference. Compared with these noise reduction methods, our algorithm is robust for various types of interference—not only random interference but also regular interference—and it does not need prior knowledge of the dynamical systems.

However, when the type of additive interference is determined in advance, the result achieved with our method is no better than a method designed for that interference. Like most other noise reduction methods, the main problem in the application of our method is the selection of different parameters. Establishing rules for choosing suitable values of parameters is one of our further research goals.

**Acknowledgments**

This work is supported by the National Natural Science Foundation of China (Grant No 60874099) and "Cognitive Computing of Visual and Auditory Information" of National Natural Science Foundation of China (Grant No 90820001).